\documentstyle[11pt,newpasp,twoside,epsf]{article}
\def\edcomment#1{\iffalse\marginpar{\raggedright\sl#1\/}\else\relax\fi}
\reversemarginpar

\begin{document}
\title{MUSICOS 1998: Optical and X-rays Observations of Flares on the RS CVn Binary HR 1099}
 \author{D. Garc\'{\i}a-Alvarez \& J.G. Doyle}
\affil{Armagh Observatory, College Hill, Armagh BT61 9DG N.Ireland, dga@star.arm.ac.uk}
\author{B.H. Foing \& J.M Oliveira\altaffilmark{1}}
\affil{Research Division, ESA Space Science Department, ESTEC/SCI-R, P.O. Box 299, 2200 AG Noordwijk, The Netherlands}
\author{D. Montes}
\affil{Departamento de Astrof\'{\i}sica, Facultad de Ciencias F\'{\i}sicas, Universidad
Complutense de Madrid, E-28040 Madrid, Spain}

\begin{abstract}
We present simultaneous and continuous observations of H$\alpha$, 
H$\beta$, Na~{\sc i} D$_{1}$, D$_{2}$ and He~{\sc i} D$_{3}$ lines
   of the chromospherically active binary HR 1099.  
   We have observed HR 1099 for more than 3 weeks almost continuously and 
monitored two flares.
   An increase in H$\alpha$, Ca~{\sc ii} H \& K, H$\beta$,  
He{\sc i} D$_{3}$  and He{\sc i} $\lambda$6678 and a
   strong filling-in of the Na{\sc i} D$_{1}$, D$_{2}$ and Mg{\sc i} b
 triplet during one of the flares are observed. 
We have found that the flares took place at the same phase
   (0.85) of the binary orbit, and both of them seems to occur near the limb. Several X-rays flares were also detected by ASM on board RXTE. Rotational modulation in the X-rays light curve has been
   detected with maximum flux when the active K1IV star is in front.
\end{abstract}

\section{Introduction}

HR 1099 is a triple system and consists of a close double-lined spectroscopic pair with a spotted and rapidly rotating K1IV primary and a
comparably inactive and slowly rotating G5V secondary in a $2\fd8$ orbit. The tertiary is a fainter K3V star $6\arcsec$ away. The system HR 1099 is one of the few RS CVn systems, along with UX Ari, II Peg, and DM UMa, that shows H$\alpha$ 
consistently in emission. Its tidally induced rapid rotation, combined with the deepened 
convection zone of a post main sequence envelope, is
   responsible for the very high chromospheric activity for its spectral 
class. \footnote{Astrophysics Group, Department of Physics, Keele University, Staffordshire ST5 5BG, UK.}

Short-period RS CVn-like systems, through their rotational modulation, could give us information about the morphology and three-dimensional
spatial distribution of stellar coronae. Possible detection of rotational modulation in the
EUV light curve of HR 1099 was reported by Drake et al. (1994), with a minimum flux occurring near the phase when
the G5 star is in front ($\phi\approx$0.5), consistent with a previously reported correlation between binary phase and X-ray intensity by
Agrawal \& Vaidya (1988).

Following previous MUSICOS campaigns on HR 1099 (Foing et al. 1994) the main goals of this campaign were to monitor for flares and to observe
chromospheric lines in order to diagnose the energetics and velocity dynamics. Also photospheric Doppler Imaging and the study of 
chromospheric activity variations were planned.

\section{Observations}
The 
spectroscopic observations have been obtained during the MUSICOS 1998 campaign (Nov-Dec),
involving OHP, SAAO, La
   Palma, Kitt Peak, ESO La Silla, Mt Stromlo, Xinglong and LNA, using both 
Echelle and Long Slit Spectrographs. A summary of the sites and instruments involved in the campaign and some of their most
important characteristics can be found in Oliveira (2001). The spectra have been extracted using the standard reduction
procedures in the IRAF package. The wavelength calibration was obtained by taking spectra of a Th-Ar lamp. The
spectra have been normalized by a low-order polynomial fit to the observed continuum. Finally, for the spectra
affected by water lines, a telluric correction have been made.

The X-rays observations were obtained with the all-sky monitor (ASM) detector on board RXTE (Levine et al. 1996). The ASM consists of three similar scanning shadow cameras, 
sensitive to X-rays in an energy band of approximately 2-12 keV, which perform sets of 90 s pointed observations ("dwells"). The analysis presented here makes use of  
light curves from individual dwell data. Light curves are available in three energy bands: 1.5-12 KeV, 3.0-5 and 5-12 keV. Between 5-10 of individual ASM
dwells of HR 1099 were observed daily by RXTE, during MUSICOS 1998 campaign. The data were binned in 1 hour intervals.

\section{Results}

HR 1099 was observed almost continuously during the campaign, approximately 3 weeks in the optical and X-rays range. 

During the campaign, two flares were observed, one at JD 2451145.51 (28-11-98) lasting about 0.63 days and a second flare at JD
2451151.07 (03-12-98) lasting about 1.1 days. This second flare shows 
an increase in the H$\alpha$ emission line, H$\beta$ and He{\sc i} D$_{3}$ lines turns into emission during
 the flare and it also shows a strong filling-in of the Na{\sc i} D$_{1}$, D$_{2}$ line (see top Fig~1). Increase in Ca~{\sc ii} H \& K
 lines, filling-in of the Mg{\sc i} b
 triplet lines and He{\sc i} $\lambda$6678 turning into emission during the second flare were also observed.

In the middle  panel of Fig~1 we plot the radial
  velocity curves of the binary system HR 1099, calculated using photospheric lines. We have also plotted the radial velocity for the two
  monitored flares, calculated using H$\alpha$. As a result, we notice that during both
 flares, the radial velocity is slightly displaced (15-20 $\mathrm{km\ s^{-1}}$) compared to the center of
 gravity of the primary. This result could be due to the fact that both flares had taken place near the limb. 

The bottom panel of Fig~1 shows the 
 $\rm{EW_{H\alpha}/EW_{H\beta}}$ ratio as a function of Julian date and as a
 function of phase. During quiescent phase we obtained values around 1 for the $\rm{EW_{H\alpha}/EW_{H\beta}}$ ratio, while during the first flare 
 we obtained values slightly bigger than in the quiescent phase (2-3). During the second flare the ratio even reaches values of 8.

In the top panel of Fig~2 we show the equivalent width as a function of Julian date and as a function of phase for several chromospheric
lines. From these results we observed that, although both flares
shown an increase in H$\alpha$ emission, the second flare produced a bigger increase. We have also observed that the first flare shows
 filling-in of H$\beta$ but for the second flare this line turns into emission. Another remarkable feature is that the Na{\sc i} D$_{1}$, 
 D$_{2}$  lines have a strong filling-in during the second
 flare. Note that both flares took place at around the same phase (0.85), but $\sim$6 days apart. Note also that we have detected rotational
 modulation of the He{\sc i} D$_{3}$ line that could be attributed to the pumping of the HeI line by coronal X-rays
 from active regions.

The long light curve of HR 1099 (bottom panel Fig~2) display evident variability on short and long time-scales. Additional to the flare at 
JD 2451151.07, several other posible flares, not observed in the optical range, can be seen (e.g., at 2451141.49 d, 2451155.21 d, 2451157.62 d and 2451159.54 d). An
additional feature of the light curve that we interpret as rotational modulation is seen around JD 2451155.21. The X-ray flux appear to peak
at $\phi\approx$0.27, after, the flux disminishes until it reaches a minimum value when the G5V star is in front of the active K1IV star ($\phi$=0.5), in agreement
with Drake et al. (1994). The
strong variability from one rotational period to other may suggest that long-term flaring is involved. These effects are observed in all the
energy bands. The flare at JD 2451151.07 shows an increase in the S band (1.5-12 keV). However, the medium energy band (3-5 keV) nor the
high energy band (5-12 keV) 
show any clear evidence for variability
suggesting this to be a soft X-rays event only. The optical flare at JD 2451145.51 was not observed in any of the X-rays bands. Spectral hardness analysis does not show clear signatures of heating and cooling.

\acknowledgments

      We wish to thanks those that have contributed to the MUSICOS 1998 campaign. Research at Armagh Observatory is grant-aided by the 
      Department of Culture, Arts and Leisure for Northern Ireland. DGA wished to thank the Space Science Department at ESTEC the finacial
      support. DM is supported by the Spanish DGESIC under grant PB97-0259. This paper made use of quick look data provided by the RXTE ASM team 
      at MIT and GSFC.

%==============================================================================
%=======================  FLARE SPECTRUM AND RADIAL VELOCITY =================

\begin{figure}[H]
\plotfiddle{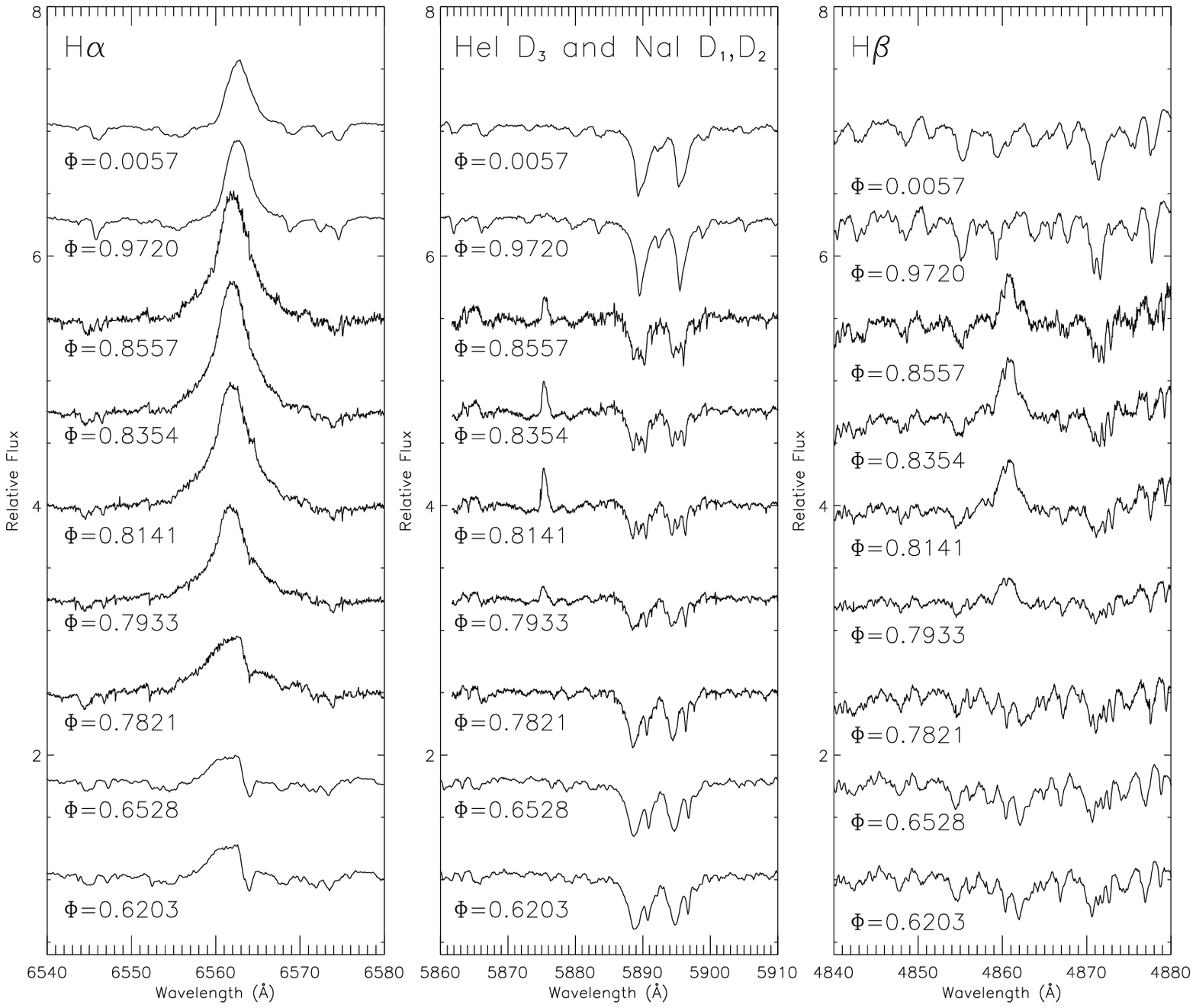}{2.6 in}{0.}{67.}{56.}{-210}{-55}
\vspace{-2.5 cm}
\end{figure}

\begin{figure}[h]
\vspace{2 cm}
\plotfiddle{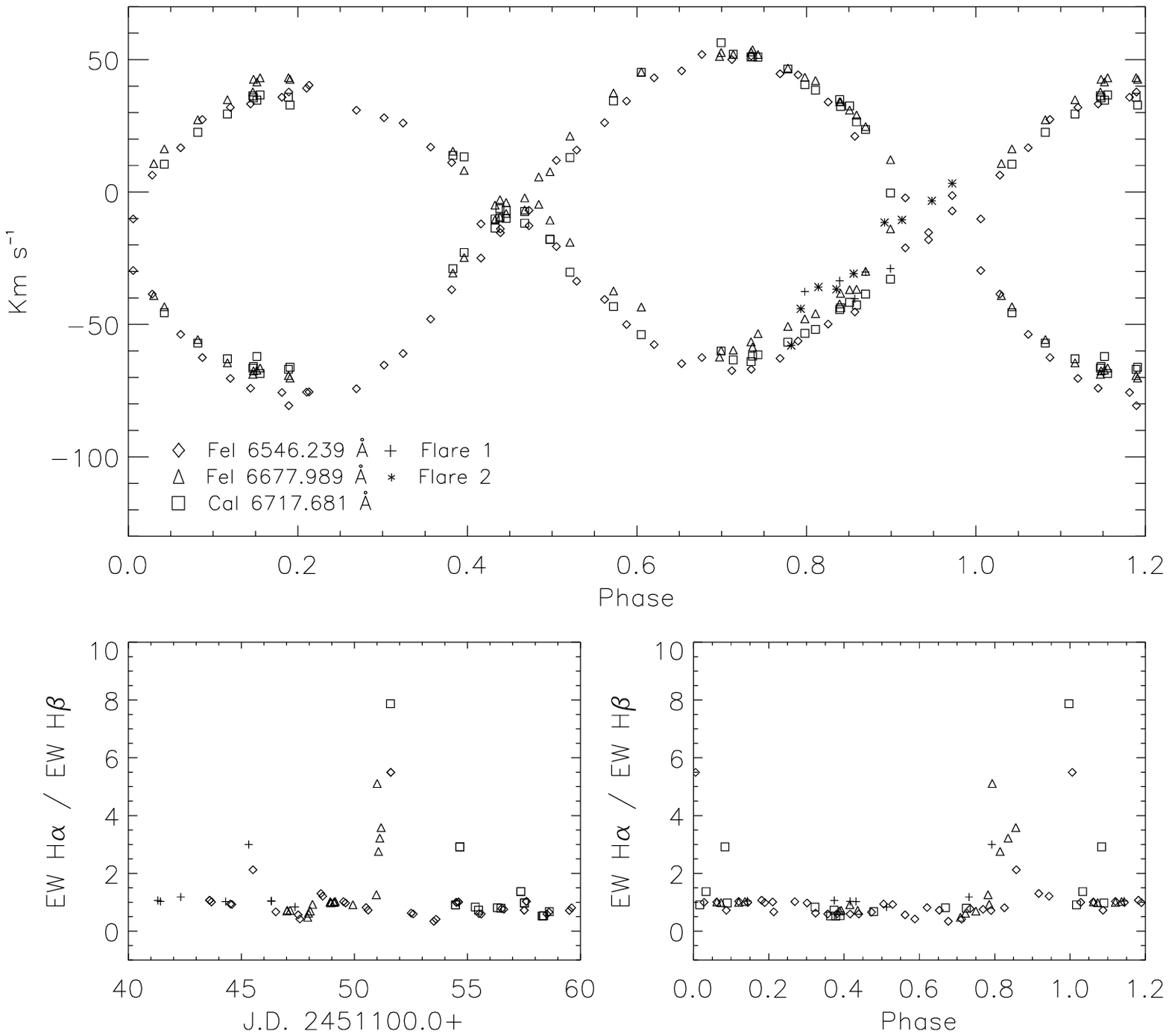}{2.6 in}{0.}{67.}{56.}{-210}{-50}

\caption{Top panels: The observed spectra for 
H$\alpha$ (left panel), He{\sc i} D$_{3}$ and Na{\sc i} D$_{1}$,D$_{2}$ (middle panel) and H$\beta$ (right panel) of the second monitored flare 
   starting at JD 2451151.07 arranged in order of the orbital phase. Middle panel: The radial velocity curves 
of the binary systems HR 1099, calculated using photospheric lines. Bottom panels: The 
 $\rm{EW_{H\alpha}/EW_{H\beta}}$ ratio as a function of Julian date and as a
 function of phase. }

\end{figure}

%==============================================================================
%=======================  EQUIVALENT WIDTH  AND X-RAY DATA =================

\begin{figure}[H]
\plotfiddle{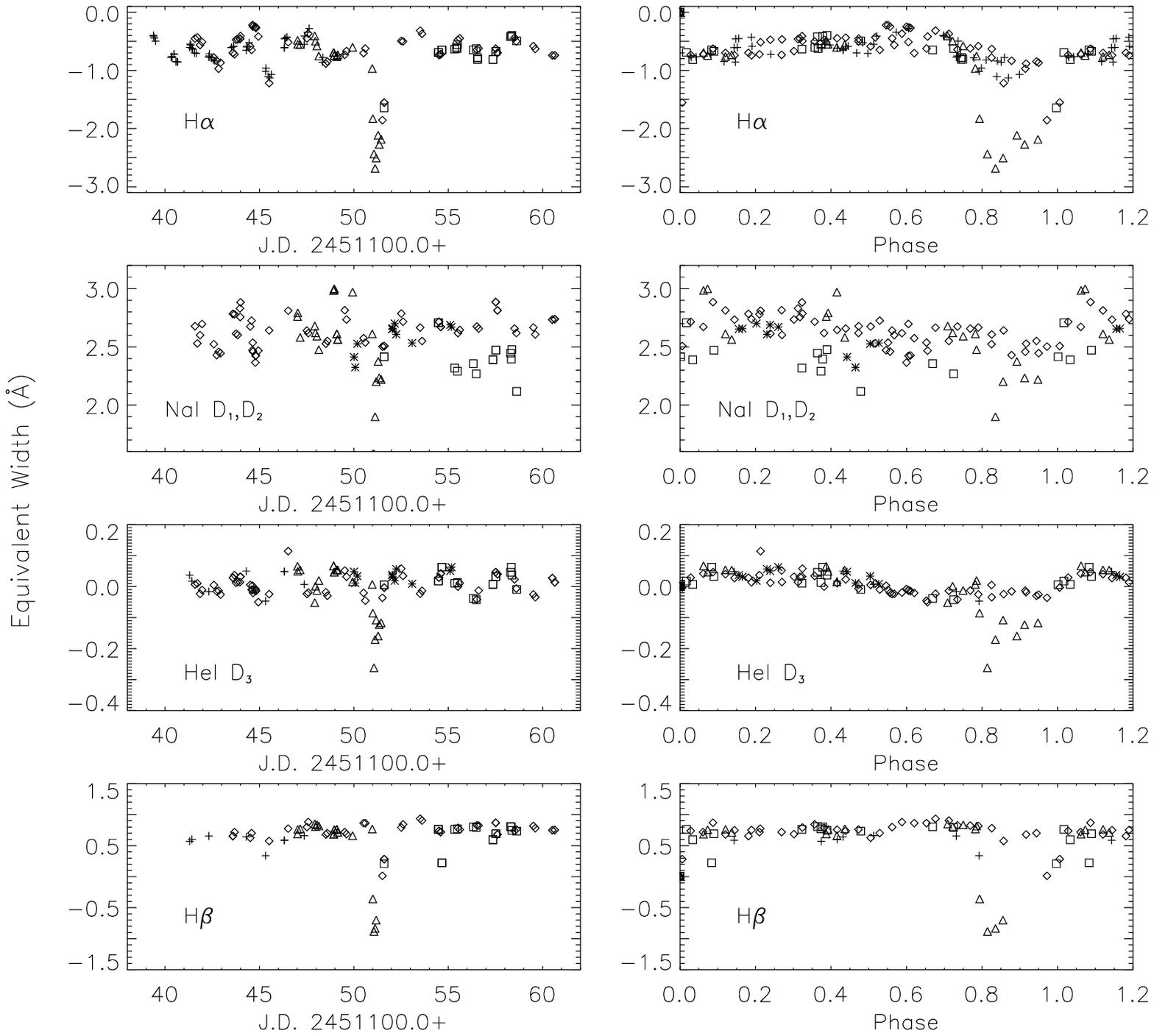}{2.6 in}{0.}{67.}{56.}{-210}{-60}
\end{figure}

\begin{figure}[h]
\plotfiddle{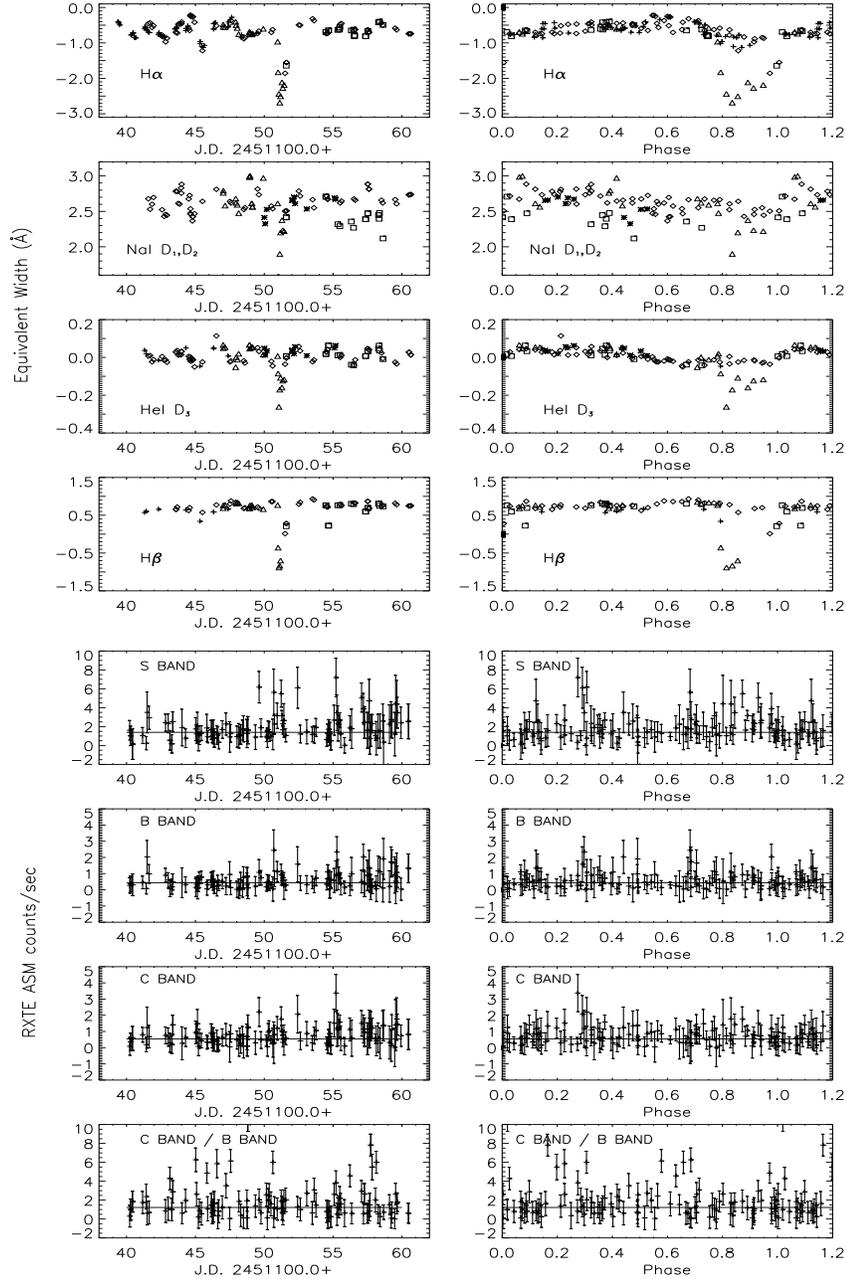}{2.6 in}{0.}{67.}{56.}{-210}{-50}
\caption{Top panels: The equivalent width as a function of Julian date and as a function of phase for H$\alpha$, Na{\sc i} D$_{1}$,D$_{2}$
, He{\sc i} D$_{3}$ and H$\beta$ lines. Bottom panels: Light curve observations of HR 1099, from the ASM instrument on RXTE satellite,
 obtained at the same time as the MUSICOS 1998 campaign. The band S (1.5-12 keV), band B (3-5 keV), band C (5-12 keV) and hardness ration C/B
 as a function of Julian date and as a function of phase are shown.}

\end{figure}

\end{document}